\documentclass{article}
\usepackage[T1]{fontenc}
\usepackage{graphicx}
\usepackage{epsf,rotate}
\usepackage{epsfig}
\usepackage{rotating}

\makeatletter

\newcommand{\LyX}{L\kern-.1667em\lower.25em\hbox{Y}\kern-.125emX\spacefactor1000}

\usepackage[T1]{fontenc}

\makeatletter

\usepackage[T1]{fontenc}
\usepackage{geometry}


\makeatother

\begin{document}

\title{Damage Spreading at the Corner Filling Transition in the 
two-dimensional Ising Model}

\author{M. Leticia Rubio Puzzo and Ezequiel V. Albano}
\date{Instituto de Investigaciones Fisicoqu\'{\i}micas Te\'{o}ricas 
y Aplicadas (INIFTA), Facultad de Ciencias Exactas, UNLP, CONICET, 
Casilla de Correo 16, Sucursal 4, (1900) La Plata, Argentina.}
\maketitle

\begin{abstract}

The propagation of damage on the square Ising lattice with a 
corner geometry is studied by means of Monte Carlo simulations. 
By imposing free boundary conditions at which competing boundary magnetic 
fields $\pm h$ are applied, the system undergoes a filling transition 
at a temperature $T_f (h)$ lower than the Onsager critical temperature $T_C$.  
The competing fields cause the formation of two magnetic domains 
with opposite orientation of the magnetization, separated by an interface 
that for T larger than $T_f (h)$ (but $T<T_C$) runs along the diagonal of 
the sample that connects the corners where the magnetic fields of different
orientation meet. Also, for $T<T_f (h)$ this interface is localized 
either close to the corner where the magnetic field is positive 
or close to the opposite one, with the same probability. 
 
It is found that, just at $T=T_f (h)$, the damage initially propagates
along the interface of the competing domains, according 
to a power law given by $D(t) \propto t^{\eta}$.

The value obtained for the dynamic exponent ($\eta^{*} = 0.89(1)$) 
is in agreement with that corresponding to the wetting transition in 
the slit geometry (Abraham Model) given by $\eta^{WT} = 0.91(1)$.
However, for later times the propagation crosses to a new regime
such as $\eta^{**} = 0.40 \pm 0.02$, which is due
to the propagation of the damage into the bulk of the magnetic domains.  
This result can be understood due to the constraints imposed 
to the propagation of damage by the corner geometry of the system
that cause healing at the corners where the interface is attached.

The critical points for the damage spreading transition ($T_D(h)$) are evaluated by extrapolation to the thermodynamic limit by using a finite-size scaling approach. Considering error bars, an overlap between the filling and the damage spreading transitions is found, such that $T_f(h)=T_D(h)$.

The probability distribution of the damage average position $P(l_0^{D})$ and that of the interface between magnetic domains of different orientation $P(l_0)$ are evaluated and compared. 
It is found that, within the nonwet phase, the average position of the damage lies slightly shifted from the interface toward the side of the largest domain. However, in the wet phase both $P(l_0^{D})$ and $P(l_0)$ are Gaussians exhibiting a single peak at the position of the diagonal of the corner sample.

\end{abstract}

\pagebreak

\section{Introduction}

The interplay between critical behavior and confinement is a challenging topic 
in the field of Condensed Matter and Statistical Physics.
In fact, the confinement of fluids, polymers, magnetic materials, etc., by 
walls that interact with the physical system, leads to the occurrence of very
 interesting phenomena such as capillary condensation, wetting, corner wetting 
 (filling), etc. \cite{dietrich, schick}. 

Particularly interesting is the case of confinement in two dimensions ($d=2$) 
close to second-order critical phenomena, due to the occurrence of strong 
fluctuations. 
Within this context the classical Ising Model has been taken as an archetypical
system for the study of capillary condensation, wetting and corner filling
by means of analytical techniques (capillary wave theories, transfer matrix 
method, etc.) and extensive numerical simulations \cite{abra1,abra2, alba1, macioleka, maciolekb, alba2} (for a review see e.g. \cite{binder2003} and references therein).

\begin{figure}
\centerline{{\epsfysize=5in \epsffile{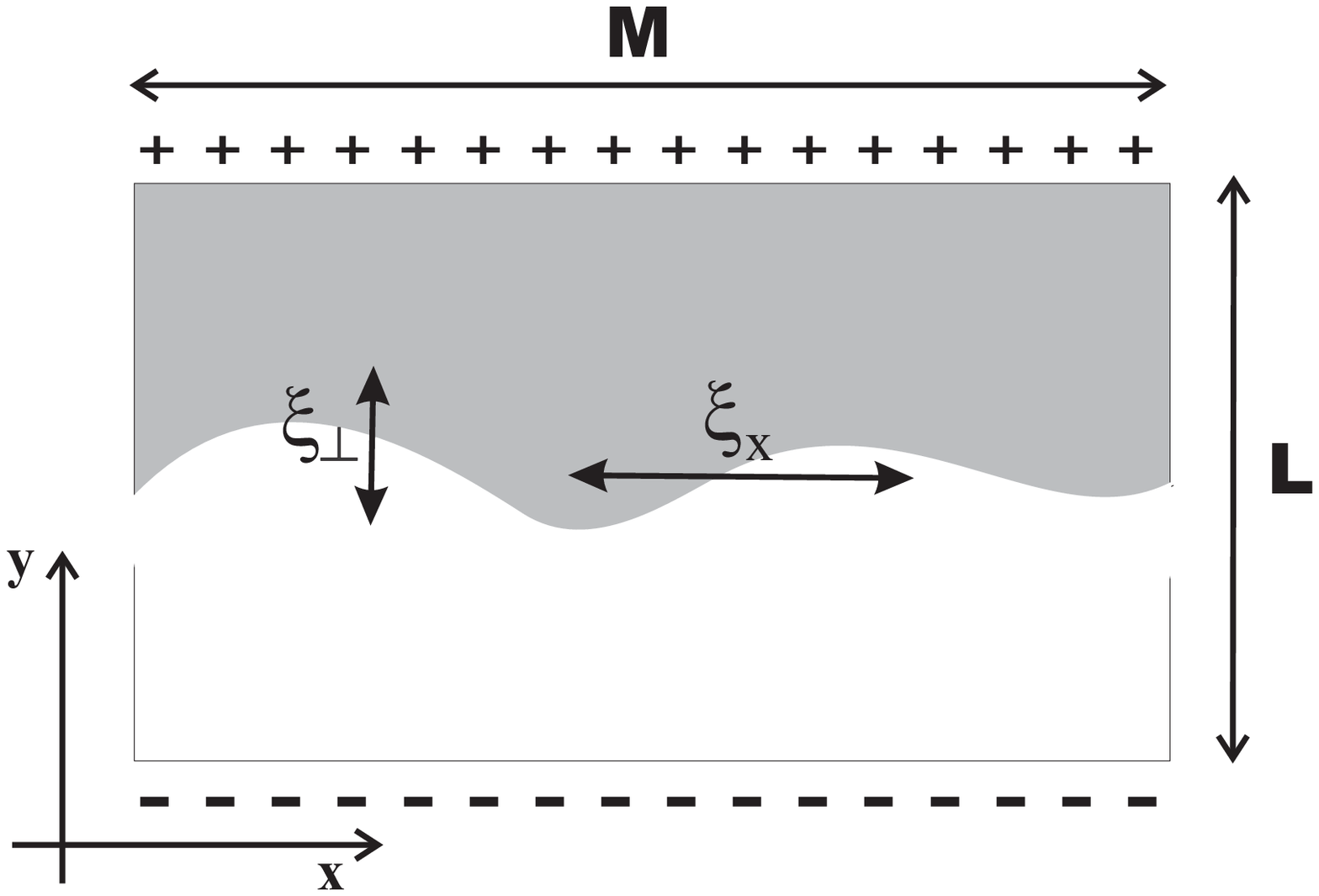}}}
\vskip 1.0 true cm
\caption{Strip geometry of size $L\times M$. The surface magnetic fields of different orientations applied at the upper (bottom) rows of the lattice are indicated by $+$ ($-$) signs. Notice that periodic (open) boundary conditions are assumed along the $x$ ($y$) axis. See notation and further details in the text.}
\label{fig1}
\end{figure}

Wetting is usually studied by using the strip geometry shown in Figure \ref{fig1}. In this case, the presence of competing magnetic fields ($h$) along the 
confinement walls induces the formation of an interface between domains where most spins are pointing up and down. 
This interface runs essentially parallel to the walls, and in a finite system, it undergoes a localization-delocalization "transition" that is the precursor 
of a true second-order wetting transition that takes place in the thermodynamic limit only ($L \rightarrow \infty$, $M \rightarrow \infty$). 
The phase diagram  (i.e. the critical curve in the $h-T$ plane, as shown 
in Figure \ref{fig3}) has been solved exactly by Abraham \cite{abra1,abra2}, 
yielding

\begin{equation}
\cosh(2h\beta) = \cosh(2K) - e^{-2K} \sinh(2K) ,
\label{eq:abra}
\end{equation}
where $J>0$ is the coupling constant, $h$ is the surface magnetic field, $\beta=1/kT$ is the Boltzmann factor, and $K=J\beta$. 

On the other hand, the study of corner filling is performed by using the geometry sketched in Figure \ref{fig2}, where the competing magnetic fields are applied at opposite corners. 
The study of this filling transition under equilibrium conditions has recently attracted growing attention \cite{duxb, cheng, napior, hauge, lipow, rejmer, parry1,pwras,glei,rascon,parry2,bed,parry3,abra3,abraM,parry5,sart,abra4,albdevir,milc,parry6,rom1,rom2,abra5,rejmer2,rascon2,giu,muller}.
Also, the filling transition upon the irreversible growth of a magnetic system has very recently been studied \cite{manias}.
In both cases, the occurrence of an interface is due to the 
presence of competing fields. 
The localization-delocalization transition of the interface in a finite system 
yields to a true second-order corner filling transition in the thermodynamic 
limit ($L \rightarrow \infty$).
The analytical expression of the equilibrium phase diagram was early conjectured by Parry et al \cite{parry3} and more recently proved rigorously by Abraham and Maciolek \cite{abraM}, yielding
\begin{equation}
\cosh(2h\beta)=\cosh(2K) - e^{-2K} \sinh^2(2K) .
\label{eq:abraMaci}
\end{equation}

The critical curve is presented in Figure \ref{fig3}, showing that for a given surface magnetic field, the filling transition takes place at a lower temperature than the wetting transition, except of course for $h=0$, where both curves converge to the Onsager critical temperature of the Ising Model ($T_C$).

In previous work we showed that the presence of interfaces between magnetic domains of different orientations, such as those observed close to the wetting transition, favors the propagation of perturbations 
in magnetic materials \cite{alrp1,alrp2,alrp3}. 
A standard technique frequently used to study the propagation of perturbations is the measurement of the Hamming distance or Damage ($D(t)$), given by
\begin{equation}
D(t)=\frac{1}{2N}\sum ^{N}_{l}\left| S^{A}_{l}(t,T)-S^{B}_{l}(t,T)\right|,
\label{eq:dam}
\end{equation}
where the summation runs over the total number of spins $N$, and the index $l\,(1\leq l\leq N)$ is the label that identifies the spins of the configurations.
$S^{A}(t,T)$ is an equilibrium configuration of the system at temperature $T$ and time $t$, while $S^{B}(t,T)$ is the perturbed configuration that is obtained from the previous one just by flipping few spins \cite{herr90, herr92}.
In order to further contribute to the understanding of the propagation of 
perturbations in magnetic materials and to clarify the role played by both the
 interfaces and the confinement geometry, the aim of this work is to report an extensive computer simulation study of damage spreading close to the corner filling transition of the $d=2$  Ising Model. 
Furthermore, the obtained results will be discussed within the context of our 
current knowledge of the corner filling transition \cite{albdevir} and compared with our previous studies of damage propagation in the strip geometry where a wetting transition is observed \cite{alrp1,alrp2,alrp3}. 
While in both cases one has fluctuating interfaces due to the presence of competing fields, the geometries used are quite different (see e.g. Figures \ref{fig1} and \ref{fig2} for the sake of comparison), so we expect that this situation will help us to contribute to the understanding of 
the effect of confinement on the properties of interfaces and the 
propagation of damage along them.

This manuscript is organized as follows: in Section II we describe the corner wetting transition and the Monte Carlo simulation method. In Section III we present and discuss the obtained results. Finally, the conclusions are stated in Section IV.

\section{Brief overview of the corner filling transition and the simulation method}

The Hamiltonian ($\cal H$) for the Ising Model in the corner geometry with competing short-range local fields at the boundaries, as sketched in Figure \ref{fig2}, is given by   

\begin{equation}
{\cal H}=-J\sum_{<i,j,m,n>} \sigma_{i,j} \sigma _{m,n} -h \sum_{i}\sigma _{i,1}- 
h\sum_{j}\sigma _{L,j}+ h\sum_{j}\sigma _{1,j}+ h \sum_{i}\sigma _{i,L},
\label{eq:ham}
\end{equation}
where $\sigma_{i,j}= \pm 1$ is the spin variable, $J>0$ is the coupling 
constant, and $h$ is the magnitude of the surface field. 
The first summation runs over all spins, 
while the remaining ones hold for spins at the surfaces where the magnetic fields are applied (see also Figure \ref{fig2}) and $h>0$ is measured in units of $J$.

It is very useful to discuss the filling  transition of a cavity by using the 
mapping between the Ising Model and a Lattice Gas. 
So, by considering the transformation spin up $\Leftrightarrow$ vapor, 
spin down $\Leftrightarrow$ liquid, respectively, one has that the 
delocalization of the interface can be thought as the growth of a 
macroscopic liquid layer wetting on the substratum.            
Let us consider a corner cavity forming an angle $\pm \phi$ with the horizontal
 axis, as shown in Figure \ref{fig2}. 
We further assume that the system is in contact with the vapor phase of the bulk at a certain temperature $T$ and chemical potential $\mu$, 
so that at coexistence one has $\mu=\mu_{sat}$.
According to (macroscopic) thermodynamic arguments it has been shown \cite{pwras}  that complete filling by the liquid is observed for $T\ge T_f$, where $T_f$ is the filling temperature given by

\begin{equation}
\theta (T_{f})=\phi,
\end{equation}
where $\theta (T_{f})$ is the contact angle of the liquid to the flat substratum. 
Consequently, one has that $T_{f} < T_{w}$, where $T_{w}$ is the critical temperature for the wetting of the plane by the liquid. 

The case studied here corresponds to $\phi=\pi /4$ (see Figure \ref{fig2}).
The corner filling transition of the Ising  Model is of second order, so one has power-law behavior of relevant quantities such that the average interface position in the y-direction ($\langle l_0 \rangle$), the parallel ($\xi_x$) and perpendicular ($\xi_{\perp}$) correlation lengths (see Figure \ref{fig2}) that describe the fluctuations of the interface in their respective directions, behave as follows

\begin{equation}
\langle l_0\rangle\propto \varepsilon^{-\beta_s}\;; \qquad \xi_{\perp}\propto \varepsilon^{-\nu_{\perp}}\;; \qquad \xi_x \propto \varepsilon^{-\nu_x},
\label{exponentes}
\end{equation}
where $\varepsilon=(h_C (T)-h)$ is the distance to the critical curve, $\beta_s=1$ is the order parameter critical exponent \cite{note}, and $\nu_{\perp}= \nu_{x}=1$ are the correlation length exponents in the direction perpendicular and parallel to the interface, respectively \cite{albdevir}.

On the other hand, the probability distribution of the position of the interface is given by
\begin{equation}
P(l_0,\varepsilon)\propto \frac{1}{\langle l_0\rangle} {\cal P} \left(\frac{l_0}{\langle l_0\rangle},\frac{\xi_{\perp}}{\langle l_0\rangle}\right),
\label{pl0}
\end{equation}
where the second scaling variable can be neglected close to criticality because both $\langle l_0\rangle$ and $\xi_{\perp}$ diverge with the same critical exponent (see equation (\ref{exponentes})).
Furthermore, at criticality and for $L \rightarrow \infty$, $\varepsilon \rightarrow 0$ one has \cite{pwras}

\begin{equation}
P(l_0)\propto \frac{1}{\langle l_0\rangle} \exp{\left(-l_0/\langle l_0\rangle\right)},
\label{Pl0crit}
\end{equation}
indicating that the probability of finding the interface close to the walls decays exponentially. 
For large enough lattices, equation (\ref{Pl0crit}) can be symmetrized so that

\begin{equation}
P(l_0)\propto \frac{1}{2\langle l_0\rangle}\left[\exp{\left(\frac{-l_0}{\langle l_0\rangle}\right)} + \exp{\left(\frac{-(L-l_0)}{\langle l_0\rangle}\right)}\right].
\label{Plosim}
\end{equation}
All these equations hold for the localized interface below but close to the critical curve shown in Figure \ref{fig3}.

On the other hand, at the critical point the distribution is expected to be flat \cite{pwras}, namely 

\begin{equation}
P(l_0;\varepsilon=0) = \frac{1}{L}, \qquad L\rightarrow \infty , 
\label{Pl0parry}
\end{equation}
so that the interface can be located at any place with the same probability.
 
Above the filling transition the average position of the interface lies along 
the diagonal of the sample with $\langle l_0\rangle=L/2$ \cite{note2}, and then  the distribution is a Gaussian given by 
\begin{equation}
P(l_0)\propto \exp{\left(-\frac{[l_0-\langle l_0\rangle]^2} {2\xi_{\perp}^2}\right)}
\label{Plogauss}
\end{equation}
with $\xi_{\perp}\propto L^\frac{1}{2}$. 
For further details on the corner filling transition see \cite{pwras} and references therein.

The spreading of damage in the corner geometry is studied by means of 
Monte Carlo simulations assuming the Glauber dynamics. 
So, a randomly selected spin is flipped with probability $p(flip)$ given 
by \cite{stan}

\begin{equation}
p(flip)=\frac{\exp (-\beta \cdot \bigtriangleup H)}{1+\exp (-\beta \cdot 
\bigtriangleup H)},
\label{eq:rates}
\end{equation}
where $\bigtriangleup H$ is the change of the Hamiltonian given by equation
(\ref{eq:ham}) due to the attempted flip and $\beta =1/kT$ is the Boltzmann 
factor. 
In order to set the time scale, we assume that during a Monte Carlo time step 
(mcs) all spins of the system ($L\times L$ in total) have the chance 
to be flipped once, on the average.                                     

\section{Results and Discussion}
                                                      
Simulations are performed on the square lattice of size $L \times L$ 
($64 \leq L\leq 2048$) by using the geometry shown in Figure \ref{fig2} and 
by assuming open boundary conditions.
Most simulations are performed by keeping the temperature constant 
($0.75 \leq T/T_C \leq 0.95$) and changing the magnitude of the surface field
($0.1 \leq h \leq 0.35$).
These ranges for the parameters are selected to avoid corrections due to the standard critical behavior of the Ising magnet that appear for $T \rightarrow T_C$ and $h \rightarrow 0$.

The dynamic behavior of the system is characterized by measuring
 the total damage or Hamming distance according to equation (\ref{eq:dam})
 and the probability distribution of the distance between the damage interface and the corner $P(l_0^{D})$ (see Figure \ref{fig2}).

In order to obtain equilibrated configurations, the ground-state configuration corresponding to $T=0$ (all spins pointing up) was annealed up to the desired set $(h,T)$ for  $10^4$ mcs. 
Subsequently, initial damage was created by flipping the spins of the diagonal perpendicular to the interface (i.e. along the vertical
 direction in Figure \ref{fig2}) according to this general rule: if the magnetization of the whole sample is positive (negative), the up (down) spins are flipped. Using this procedure, it is assured that the initial damage is always $D(t=0) \leq L/2$ \cite{note2}. This kind of perturbation reproduces the effect of a large magnetic field applied at the diagonal of the sample and pointing away to the direction opposite to the total magnetization.

Subsequently, the time evolution of the damage was recorded.
Figure \ref{fig4} shows plots of $D(t)$ versus $t$ obtained by keeping $T=0.80 T_C$ constant and varying the surface field $h$. 
The observed behavior resembles the results, already published, 
corresponding to the wetting transition in a strip geometry \cite{alrp3}.
In fact, the short-time behavior of the curves involves the healing of the damage initially created in the bulk of the domains that takes place up to $t \sim 50$ mcs. Subsequently, the propagation of the damage along the interface becomes relevant. 
While for weak fields one observes damage healing (e.g. $h \leq 0.35$), by 
increasing the field the damage starts to propagate monotonically and 
a power-law behavior of the form 

\begin{equation}
D(t)\propto t^{\eta},
\label{eq:pwlaw}
\end{equation}
where $\eta$ is an exponent, can be proposed as in the case of the strip geometry \cite{alrp3}.

The onset of a power-law divergence, for a sample of size $L$, is considered as the signature for a size-dependent ``critical'' behavior of the damage process. In fact, as shown in Figure \ref{fig5inset}, the power-law behavior is observed at slightly different surface magnetic fields when the size of the lattice is changed.
Moreover, simulations performed using lattices of different size confirm 
that the power-law behavior of the damage becomes more evident for 
larger samples, as shown in Figure \ref{fig5inset}. 
These results anticipate that one would need to perform a proper extrapolation of the data in order to obtain the actual critical points in the thermodynamic limit. Before tackling this issue, let us first discuss the behavior of $D(t)$ as shown in Figure \ref{fig5inset}.
An overview inspection reveals that three different 
time regimes can easily be distinguished, as follows:

1) During the short-time regime ($t<t_{min}$), the damage decreases monotonically reaching an absolute minimum at $t_{min}$, which only depends slightly on the lattice size (see dashed-dotted line in Figure \ref{fig5inset}). 
This behavior can be understood by recalling that the damage is 
initially generated along a diagonal line of the sample, in the 
direction perpendicular to the spin up-spin down 
interface (see Figure \ref{fig2}). 
This initial condition creates - almost linear - damaged regions in the bulk of well-ordered magnetic domains (see the snapshot of Figure \ref{fig6} taken for $t=1$mcs). 
Due to the very low density of broken bonds in the bulk, the damage cannot propagate and becomes quickly healed (see the snapshot of Figure \ref{fig6} taken for $t=5$mcs) while a small fraction of the initially created damage still remains at the interface (see the snapshot of Figure \ref{fig6} taken for $t=28$mcs).
Since the initial damage is distributed along the line one has $D(t=0)\sim L^{-1}$, in agreement with the equispaced intercepts shown in Figure  \ref{fig5inset}  for $t=0$.
However, at $t_{min}$ one has only few damaged sites along the center of the interface, so one expects $D(t_{min})\sim L^{-2}$, a result that is corroborated by the numerical data as shown in the inset of Figure \ref{fig5inset}.  
Summing up, the initial decrease of the damage is due to damage healing in the bulk.

After reaching $t_{min}$ one observes the growth of the damage. 
This behavior can be better observed in the scaled plot of 
Figure \ref{fig5scaling} that attempts to collapse all curves 
corresponding to different lattices. 
Recall that a perfect collapse cannot be achieved not only due to the existence of prefactors but also because $D(t=0)\sim L^{-1}$ while $D(t_{min})\sim L^{-2}$.
However, the scaling plot of Figure \ref{fig5scaling} is useful in order to clearly show the power-law divergence of the damage above $t_{min}$.
Here, two well-defined regions can also be distinguished (recall that these regions become also evident in Figure \ref{fig5inset}).

2) For $t>t_{min}$ the damage increases according to equation (\ref{eq:pwlaw}) and the best fit for the data gives $\eta^*=0.89(1)$ (see the dashed line in Figure \ref{fig5scaling}).
It is worth mentioning that this exponent is in good agreement with our previous measurement of damage spreading along the interface in the strip geometry (see Figure \ref{fig1}) at the wetting transition, which yields $\eta^{WT}=0.91(1)$ \cite{alrp3}. 
So, we conclude that the critical increase of the damage, which takes place for $t_{min}<t<t_{cross}$, where $t_{cross}$ is the crossover time to a third regime that will be discussed below, is due to the propagation of damage along the spin up-spin down interface characteristic of the filling transition.
This statement is further confirmed by the snapshot configuration shown in Figure \ref{fig6} for $t=500$mcs.
Then, one has that the exponent describing the propagation of the damage along the interface between magnetic domains of opposite magnetization is given by $\eta^I=0.90(2)$, where the error bars are large enough to account for the fact that $\eta^I=\eta^{WT}=\eta^*$.
Our findings are also consistent with the fact that the propagation of damage along the interface may be independent of the nature of the studied phenomena - wetting or filling - provided that the measurements are performed at criticality.

3) Let us now discuss the propagation of damage after the crossover time $t_{cross}$. Here, a power law is also observed (see the dotted line in Figure \ref{fig5scaling}) but the best fit of the data according to equation (\ref{eq:pwlaw}) yields $\eta^{**}=0.40(2)$. 
This result, measured at the critical point, depends neither on $L$ nor on $h_D(L)$. 
So, we conclude that this exponent characterizes the propagation of the damage from the interface into the bulk of the corner geometry (see the snapshot of Figure \ref{fig6} taken for $t=50000$mcs). 
At this latter stage the propagation of the damage along the interface 
has ceased due to the constraint imposed by the corner geometry and 
one observes a slower propagation ($\eta^{I}> \eta^{**}$) into the 
bulk adjacent to the interface. 
Also notice that the damage generated along the interface and close to it follows the fluctuations of the actual position of the interface.

Of course, during the second time regime ($t_{min}<t<t_{cross}$) the damage does not strictly propagate along the interface only, since one also expects the onset of propagation in the direction perpendicular to it. 
However, due to the fact that the magnitudes of the exponents are quite different, this effect is no longer observed in the simulations.

After discussing the dynamics of the damage, we will like to focus our 
attention on the location of the damage spreading critical points 
in order to draw the corresponding phase diagram. 
For this purpose, note that the best fit of the power-law 
behavior according to 
equation (\ref{eq:pwlaw}) - within the long-time behavior for $t>t_{cross}$ - allows us to identify the size-dependent critical field ($h_D(L)$) for a given 
temperature (see e.g. Figure \ref{fig5inset} that corresponds to $T=0.80T_C$).
It is well known that in numerical simulations, performed by using finite samples, "critical" points are shifted and rounded due to finite size effects. 
So, as in the case of the wetting transition \cite{alba2}, we propose the following Ansatz for the shifting of the damage-healing critical point

\begin{equation}
h_{D}(L) -  h_{D}(\infty) \sim L^{-\gamma} , \qquad  \mbox{T=constant} 
\label{eq:hfinsize}
\end{equation}
where $\gamma$ is an exponent and $h_{D}(\infty)$ is the true critical field in the thermodynamic limit.

The best fits for the data corresponding to four different temperatures, as shown in Figure \ref{fig7}, were obtained by taking $\gamma=2$. 
The extrapolated values of the critical field are listed in Table 1 
that also includes, for the sake of comparison, 
the critical fields obtained by solving the exact solution of the corner-filling phase diagram \cite{abra1, abra2} given by equation (\ref{eq:abraMaci}).

\vspace{0.3cm}
{\centering \begin{tabular}{|c|c|c|}
\hline 
$T/T_C$&
$h_D(\infty)$ eq. (\ref{eq:hfinsize})&
$h_f(\infty)$ eq. (\ref{eq:abra}) \\
\hline 
\hline 
0.75&
$0.400 \pm 0.005$&
0.3975\\
\hline 
0.80&
$0.360\pm 0.005$&
0.3505\\
\hline 
0.90&
$0.245\pm 0.005$&
0.2411\\
\hline 
0.95&
$0.175\pm 0.005$&
0.1683\\
\hline 
\end{tabular}\par}

{\bf Table I.} Critical field for damage spreading ($h_{D}(\infty)$) obtained by extrapolation to the thermodynamic limit with the aid of equation (\ref{eq:hfinsize}). The third column corresponds to exact values of the critical field correspondig to the filling transition as obtained by using equation (\ref{eq:abraMaci}).
\vspace{0.3cm}

As follows from Table 1 and Figure \ref{fig3}, the critical points for damage spreading in the corner geometry are indistinguishable from those of the exact solution of the filling transition (within error bars). 
This result is in contrast with our previous studies close to the wetting transition in a strip geometry where one has $h_{D}(\infty)<h_{w}(\infty)$
(see also Figure \ref{fig3}) and the damage spreading transition is located in 
the nonwet phase of the phase diagram. 
We expect that this behavior may be due to geometrical constraints imposed 
by the corner array. 
In fact, in the strip geometry and for $h < h_{w}(\infty)$ one has that the interface is still bound to one of the walls.  
However, even in this nonwet phase but close to the wetting transition, 
the damage propagates along the interface without geometrical constraints.
 On the other hand, in the corner geometry when the interface is bound 
to one corner, the spatial propagation of the damage is restricted.

In order to gain further insight into the spatiotemporal propagation of 
the damage we also measured the probability distribution of the 
 distance from the damage zone to the corner ($P(l_0^{D})$).
The distribution was evaluated along the diagonal of the sample as shown in Figure \ref{fig2}. 
Figure \ref{fig8} shows a summary of the obtained results.

For $h\ll h_f(\infty)$ (e.g. $h=0.20$) the distribution is almost flat with 
two small peaks close to the corners. 
Approaching the transition by increasing the field these
 peaks develop and become slightly shifted toward the center of the sample 
(e.g. $h=0.21$ and $h=0.22$ in Figure \ref{fig8}).
This double-peaked structure indicates that the 
damage remains bound to each corner with the same probability as expected for the case of the nonwet phase.
On the other hand, for $h\sim h_f(\infty)$ (e.g. $h=0.23, 0.24$ in Figure \ref{fig8}) the distribution becomes a Gaussian centered along the middle of the sample. 
The Gaussian structure of $P(l_0^{D})$ remains even for $h\gg h_f(\infty)$ 
(e.g. $h=0.30$ in Figure \ref{fig8}).

It is also very useful to compare $P(l_0^{D})$ with the probability 
distribution of the position of the interface between domains of opposite 
magnetization $P(l_0)$. 
Figure \ref{fig9} shows plots of both $P(l_0^{D})$ and $P(l_0)$ obtained far below the filling transition.
One observes that for the domain interface $P(l_0)$ decays exponentially, 
as expected according to equation (\ref{Plosim}), indicating that for this set 
$(h,T)$ such an interface is strongly localized close to each corner with the same probability. 
On the other hand, while the interface of the damage is still bound, 
its average position lies further apart from each corner, 
as evidenced by the double-peaked structure of $P(l_0^{D})$ (see Figure \ref{fig9}).
So, Figure \ref{fig9} provides clear evidence that the damage is located in the neighborhood of the interface between domains but slightly shifted toward
 the bulk of the sample, or more specifically within the largest domain.

Close to the filling transition (Figure \ref{fig10}) one has that the interface between domains exhibits a flat distribution, as expected from equation (\ref{Pl0parry}), indicating that the interface could be found at any place of the sample. 
However, $P(l_0^{D})$ exhibits a single peak centered at the middle of 
the sample reflecting the inertia of the damage in order to follow the displacement of the interface. 
Furthermore, the abrupt decay of the damage close to the corners, which is evidenced 
in both Figures \ref{fig9} and \ref{fig10}, is due to the healing effect caused by the neighboring magnetic fields. 
This effect prevents the damage from approaching the corners acting as an effective 
repulsive effect.
Also, this result is consistent with our previous conclusion that the damage is essentially located within the largest magnetic domain.

Finally, within the wet phase both $P(l_0^{D})$ and $P(l_0)$ are centered around the diagonal of the sample (see Figure \ref{fig11}) in agreement with the fact that the interface is delocalized .

The main conclusions that follow from the comparison between $P(l_0^{D})$ and $P(l_0)$ (see Figures \ref{fig9}, \ref{fig10}, \ref{fig11}) are, on the one hand, the operation of an effective repulsion of the damage at the corners where the fields have the same sign, and, on the other hand, that the damage becomes attached to the interface but preferentially located toward the largest domain. 
These results are in contrast to previous observations performed by studying the spreading of the damage close to the wetting transition using the strip
 geometry where the damage was uniformly and symmetrically distributed along the interface  between competing domains \cite{alrp3}.
We expect that this difference between corner and strip geometries may explain 
the fact that for the former the phase diagram for damage spreading 
matches that of corner filling (Figure \ref{fig3}) while for the latter the damage critical points lie within the nonwet phase (Figure \ref{fig3}). 

Let us now recall that, within the wet phase, the Gaussian distribution of the 
interface profile (equation (\ref{Plogauss})) translates into a Gaussian 
distribution of the magnetization $m$ given by \cite{albdevir} 

\begin{equation}
P_{L}(m)\propto \exp(-\frac{m^{2}L^{2}\beta}{2\chi_L})
\label{Pm}
\end{equation}
where $\chi_L$ is the magnetic susceptibility and $\beta=1/kT$ is the Boltzmann factor.
Equation (\ref{Pm}) simply reflects the fact that the interface is located, 
on average, along the diagonal of the sample, the magnetic domains 
of different magnetization having the same average size. 
On the other hand, the susceptibility diverges with the lattice size according to $ \chi_{L} \propto \langle M^2 \rangle \propto L$ \cite{albdevir}.
In view of these facts we also measured the width of the Gaussian distributions $P(l_0^{D})$ (see the full line in Figure \ref{fig11}) within the wet phase, given by  $\langle s^{2}_{D} \rangle$.
The inset of Figure \ref{fig12} shows log-log plots of  $\langle s^{2}_{D} \rangle$ versus $[h-h_f]$ obtained by keeping $T=0.90 T_C$ constant and for different lattice sizes.
A preliminary inspection shows that $\langle s^{2}_{D} \rangle$ increases when the field approaches the critical value, as well as when larger lattices are considered, suggesting that $\langle s^{2}_{D} \rangle \sim \chi_L$, within the wet phase. In order to test this observation, it is worth mentioning that starting from the distribution of the magnetization (equation (\ref{Pm})) it is possible to obtain the standard relationship between the susceptibility and the fluctuations of the order parameter given by  \cite{albdevir}

\begin{equation}
k_B T \chi = L^2 (\langle m^2 \rangle - \langle |m| \rangle ^2)= L^2 \cdot \tilde{\chi} (L \varepsilon),
\label{chi}
\end{equation}
where $|m|$ is the absolute value of the magnetization, $\tilde{\chi} (z)$ is a scaling function, and $\varepsilon = (h_f - h)$. 
On the other hand, within the wet phase, one has $\varepsilon < 0$ and  $\langle |m| \rangle =0$ (at least in the thermodynamic limit). 
Then, if the proposed proportionality ($\langle s^{2}_{D} \rangle \sim \chi_L$) holds, one would have
\begin{equation}
\langle s^{2}_{D} \rangle \sim  L^2 \cdot \tilde{f} (L \varepsilon),
\label{s2}
\end{equation}
where $\tilde{f}$ is a scaling function.
Figure \ref{fig12} shows the corresponding scaling plot ($s^{2}_{D} \rangle  L^{-2}$ versus $L (h - h_f)$) as obtained by using the data of the main panel. 
By considering the involved errors and the relatively small samples that can be simulated with our computational resources, we conclude that the collapse is acceptable and most likely equation (\ref{s2}) should hold.

\section{Conclusions}

\begin{figure}
\centerline{{\epsfysize=5in \epsffile{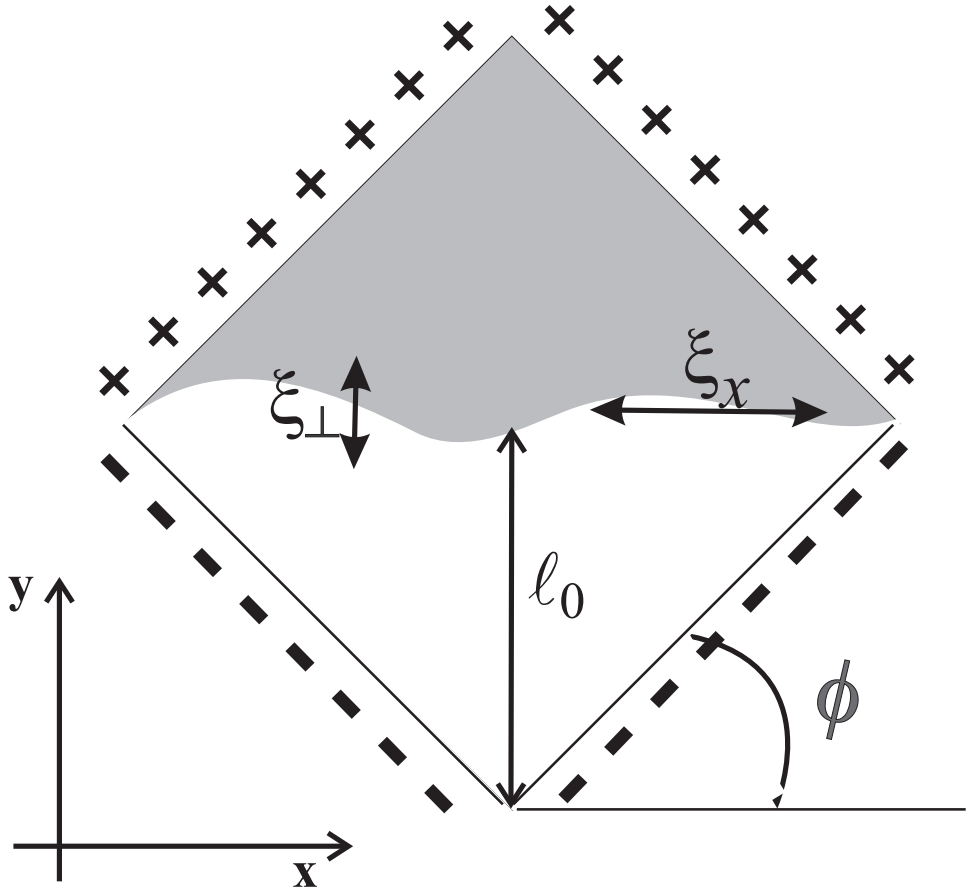}}}
\caption{Corner geometry of size $L \times L$. The signs $+$ and $-$ indicate the surfaces where the competing surface magnetic fields are applied. In this case, the boundary conditions are open for all sides of the sample. See notation and further details in the text.}
\label{fig2}
\end{figure}

\begin{figure}
\centerline{{\epsfysize=5in \epsffile{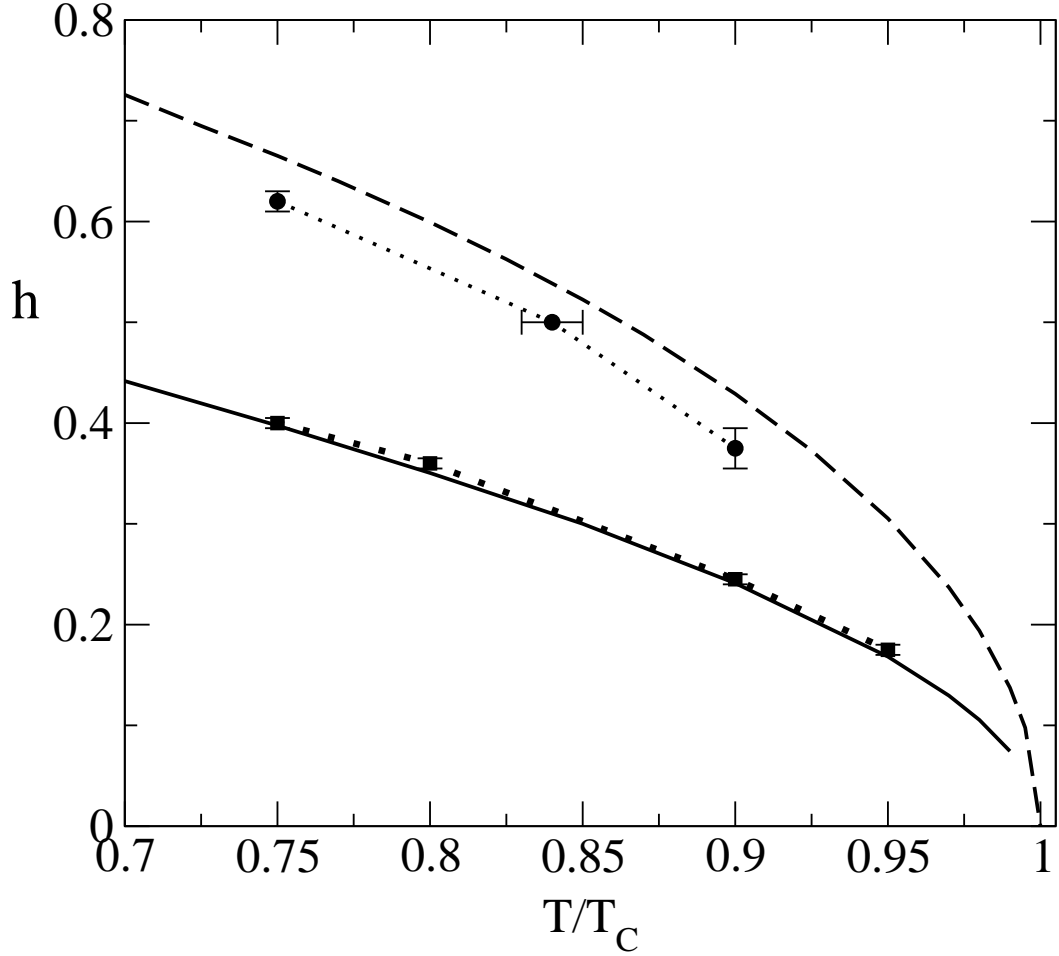}}}
\vskip 1.0 true cm
\caption{Phase diagram of $h$ versus $T/T_{C}$. The dashed line corresponds to the wetting transition (equation (\ref{eq:abra})) while the full line corresponds to the corner filling transition (equation (\ref{eq:abraMaci})).
The circles are obtained for damage transition at the strip geometry \cite{alrp3} and are shown for the sake of comparison. The squares correspond to the results obtained in the present work for the damage spreading transition at the corner geometry.}
\label{fig3}
\end{figure}

\begin{figure}
\centerline{{\epsfysize=5in \epsffile{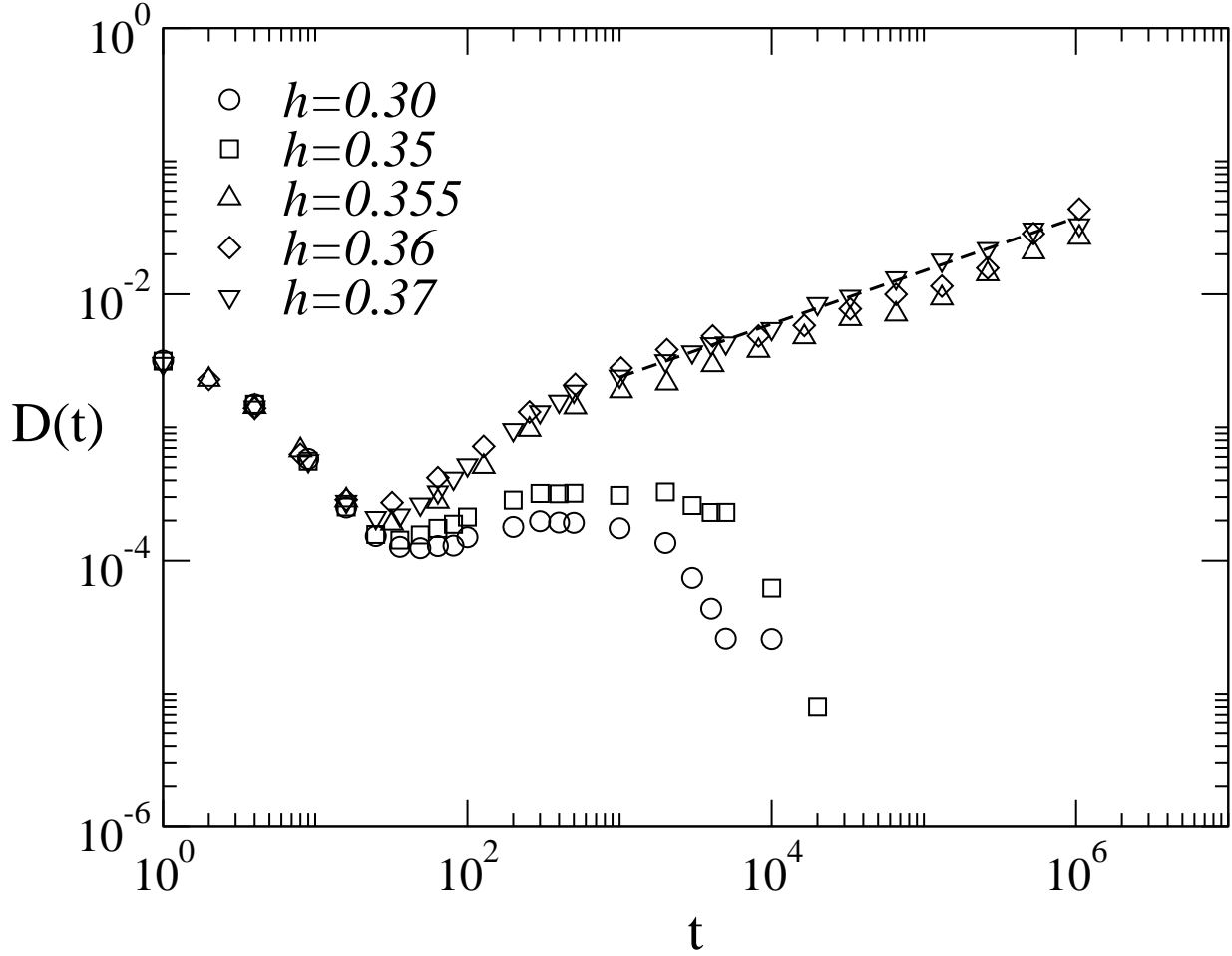}}}
\vskip 1.0 true cm
\caption{Log-log plot of damage ($D(t)$) as a function of time. Results obtained for different values of the surface magnetic fields $h$, as listed in the figure. The simulations were performed by using a lattice of size $L=256$ and for $T=0.80T_{C}$. The dashed line has slope $\eta^{**} = 0.40$.}
\label{fig4}
\end{figure}

\begin{figure}
\centerline{{\epsfysize=5in \epsffile{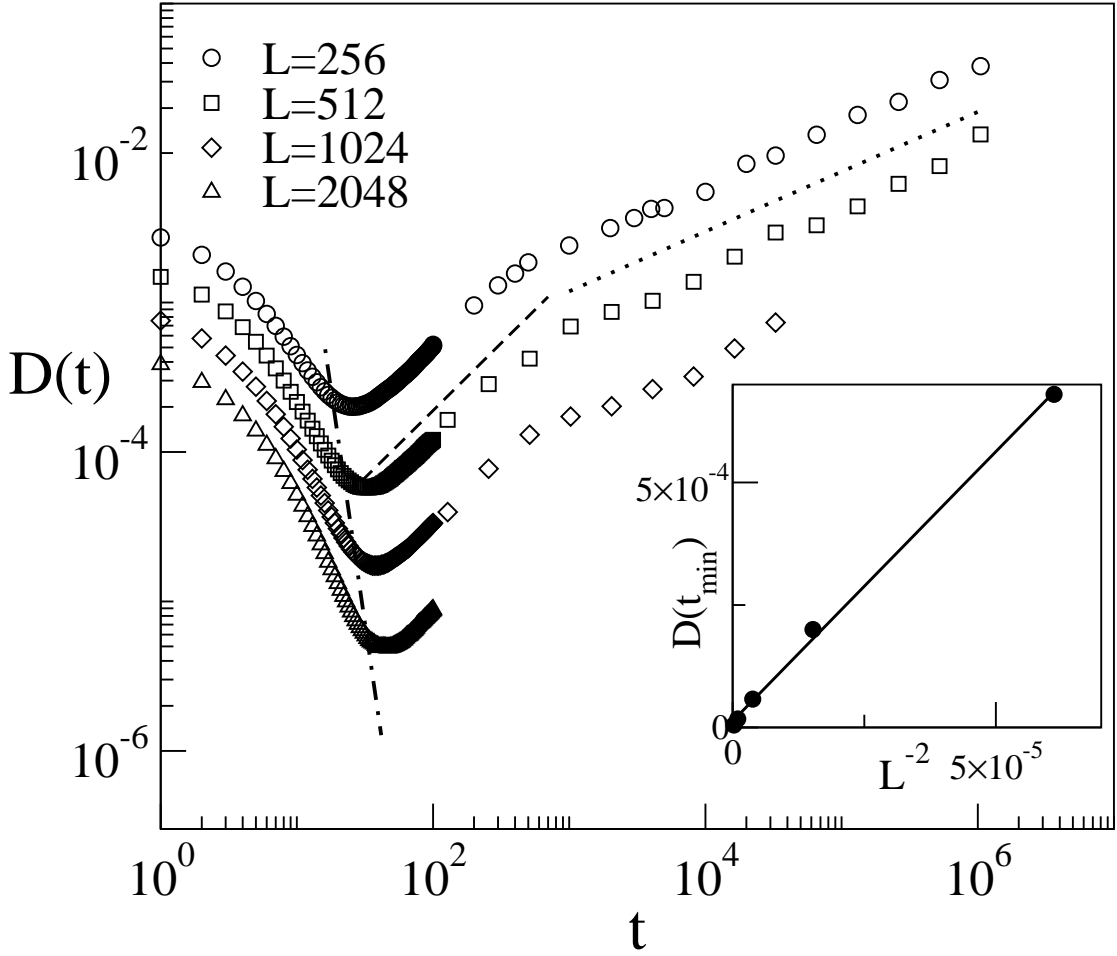}}}
\vskip 1.0 true cm
\caption{Log-log plot of $D(t)$ versus $t$, obtained for $T=0.80 T_C$ and at the size-dependent ``critical'' magnetic field $h_D(L)$: $h_D(L=256)=0.355$, $h_D(L=512)=0.3675$, $h_D(L=1024)=0.37$, and $h_D(L=2048)=0.38$.
The dashed line has slope $\eta^{*} = 0.89$ and the dotted line has slope $\eta^{**} = 0.40$. The dashed-dotted line shows the slight dependence of the minimun value of the damage ($D(t_{min})$) on L. The inset shows a linear-linear plot of $D(t_{min})$ versus $L^{-2}$.}
\label{fig5inset}
\end{figure}

\begin{figure}
\centerline{{\epsfysize=5in \epsffile{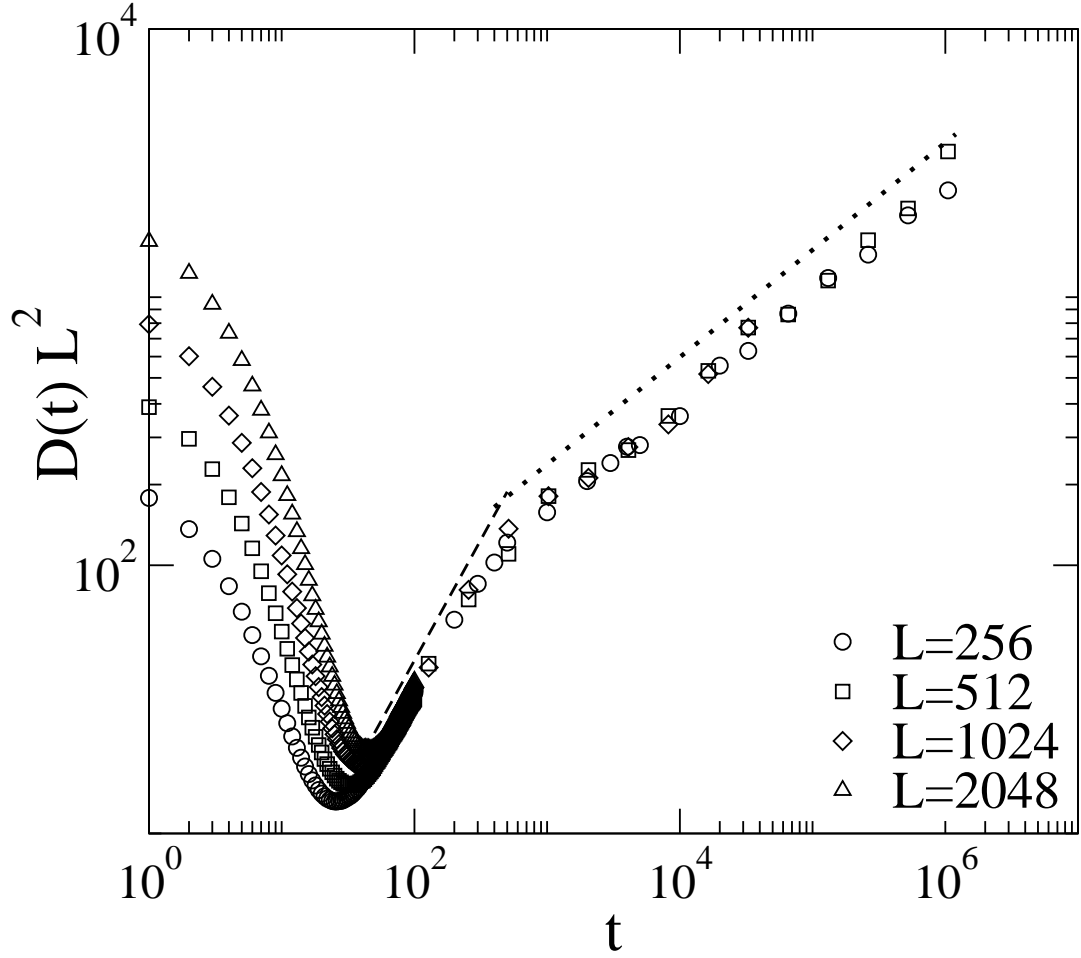}}}
\vskip 1.0 true cm
\caption{Scaling plot ($D(t)\cdot L^2$ versus $t$) of the data shown in Figure \ref{fig5inset}. The departure of the data observed for $t \rightarrow 0$ is consistent with the initial condition involving the generation of a damaged line with $D(0) \sim L^{-1}$. The dashed line has slope $\eta^{*} = 0.89$ and the dotted line has slope $\eta^{**} = 0.40$.}
\label{fig5scaling}
\end{figure}

\begin{figure}
\centerline{{\epsfysize=7in \epsffile{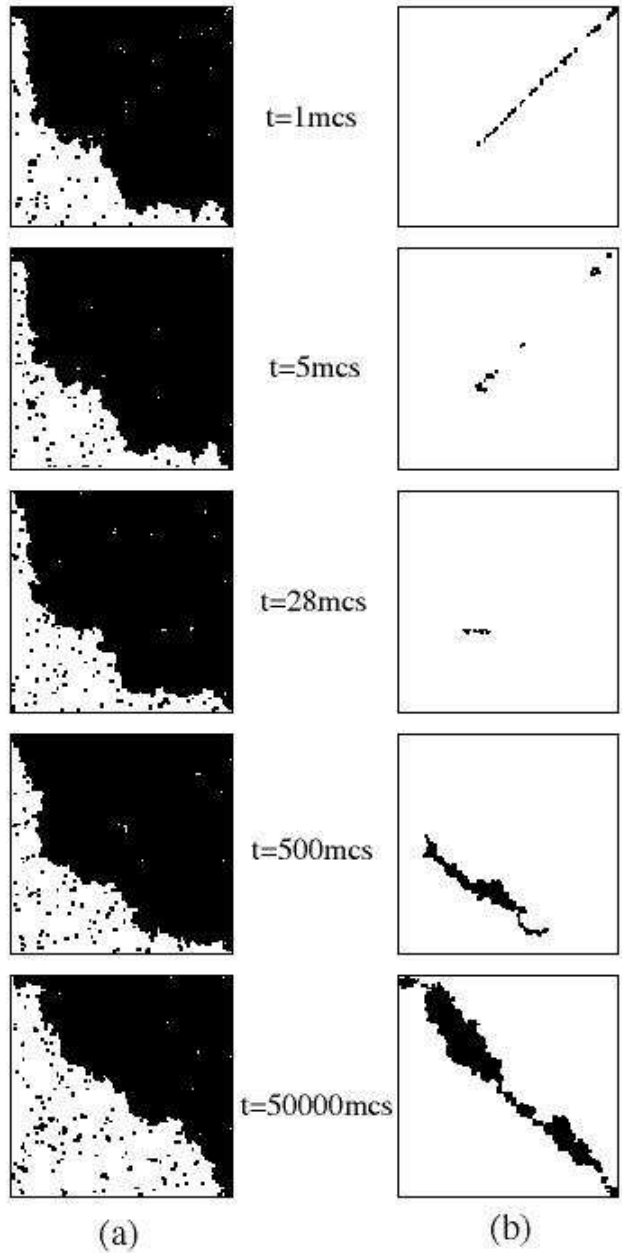}}}	
\vskip 1.0 true cm
\caption{(a) Snapshots of equilibrium configurations obtained at $T=0.8T_C$, $h=0.37$, $L=256$, and different times from top to bottom: $t=1$ mcs, $t=5$ mcs, $t=28$ mcs, $t=500$ mcs, and $t=50000$ mcs.
  (b) Snapshot pictures of the damaged sites obtained from the lattices shown in  (a) by applying linear damage along the diagonal. Damaged sites are shown in black and undamaged sites are left white.}
\label{fig6}
\end{figure}

\begin{figure}
\centerline{{\epsfysize=5in \epsffile{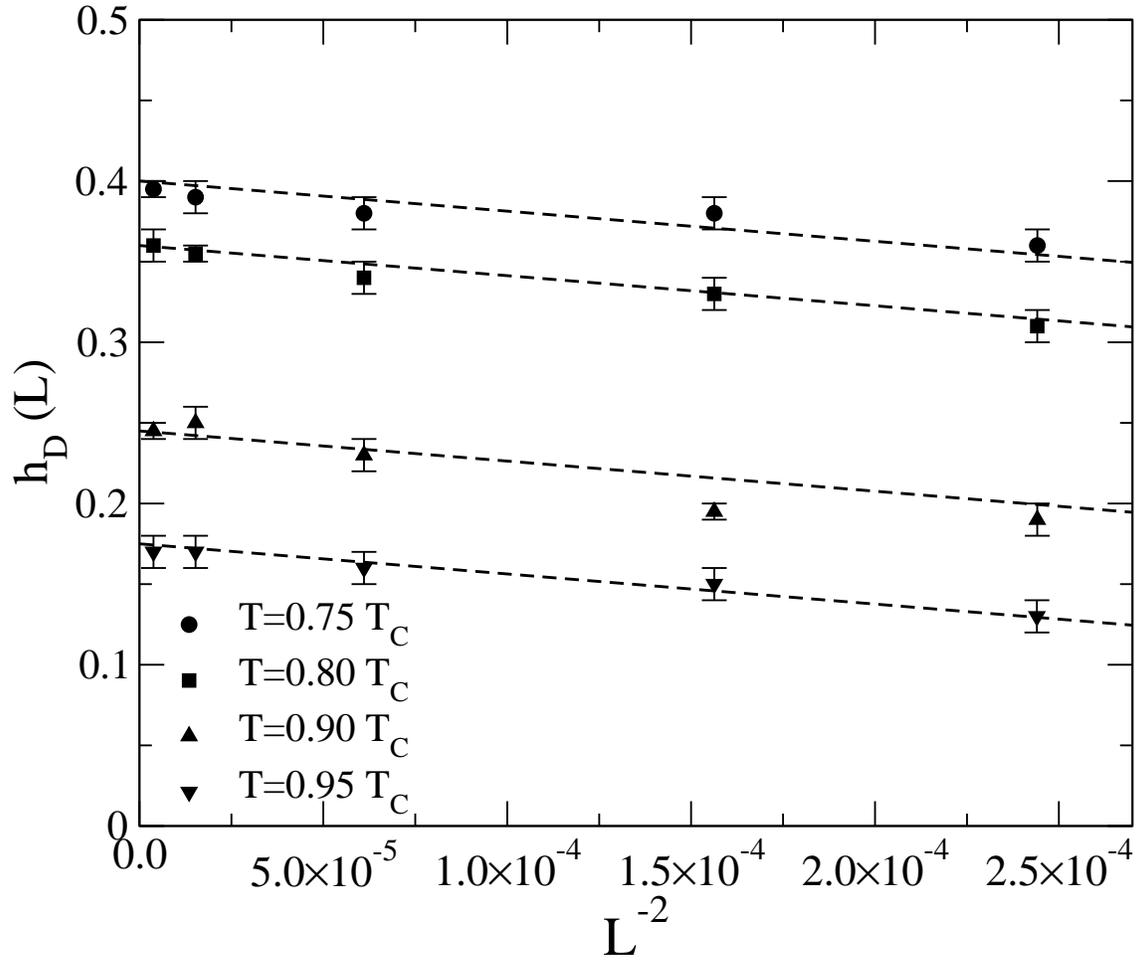}}}
\vskip 1.0 true cm
\caption{Plots of the size-dependent ``critical'' magnetic fields of the damage transition $h_{D}(L)$ as a function of $L^{-2}$, as obtained for the temperatures listed in the Figure.}
\label{fig7}
\end{figure}

\begin{figure}
\centerline{{\epsfysize=5in \epsffile{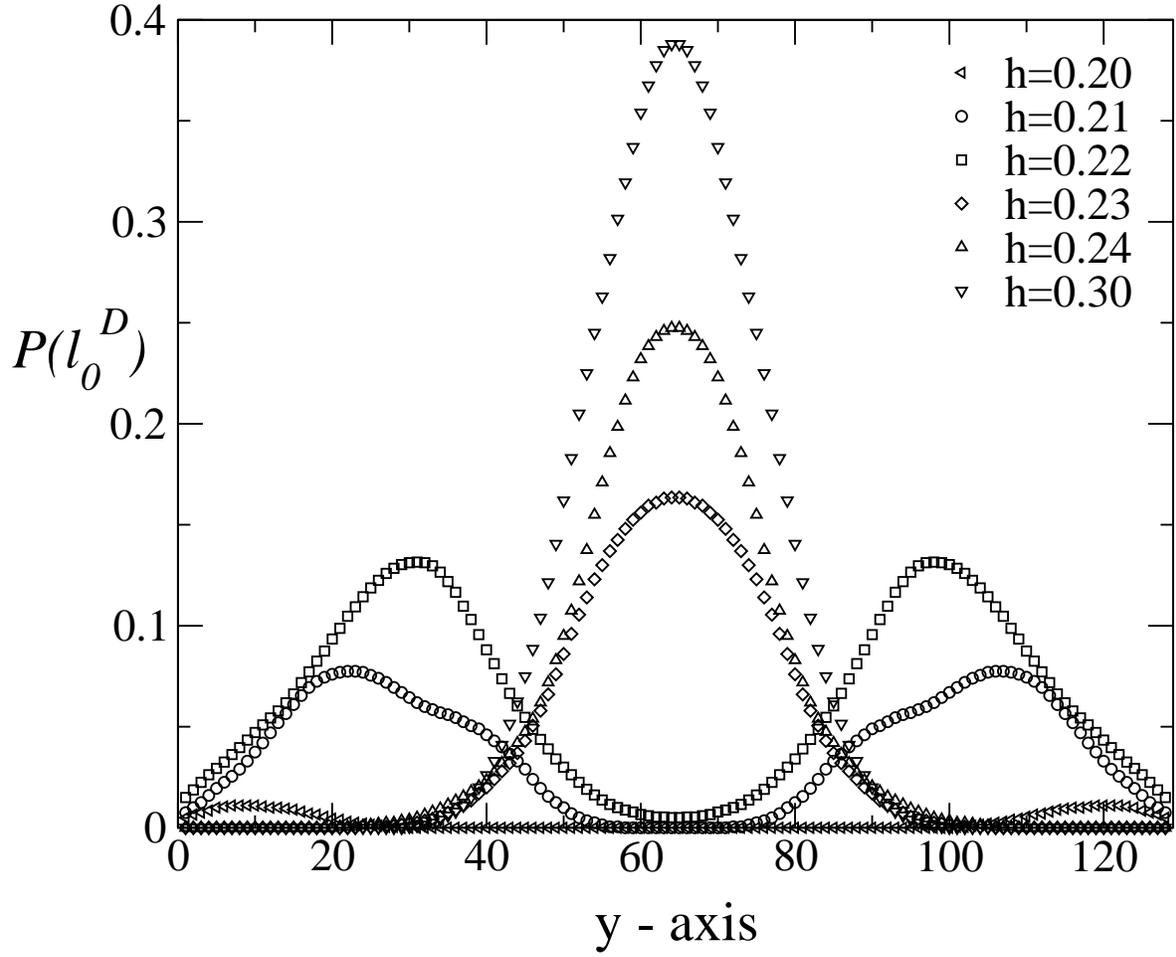}}}
\vskip 1.0 true cm
\caption{Plot of probability distribution of the position of the damage $P(l_0^D)$, obtained for $L=128$, $T=0.90 T_C$, and different values of surface magnetic field $h$, as listed in the figure.}
\label{fig8}
\end{figure}

\begin{figure}
\centerline{{\epsfysize=5in \epsffile{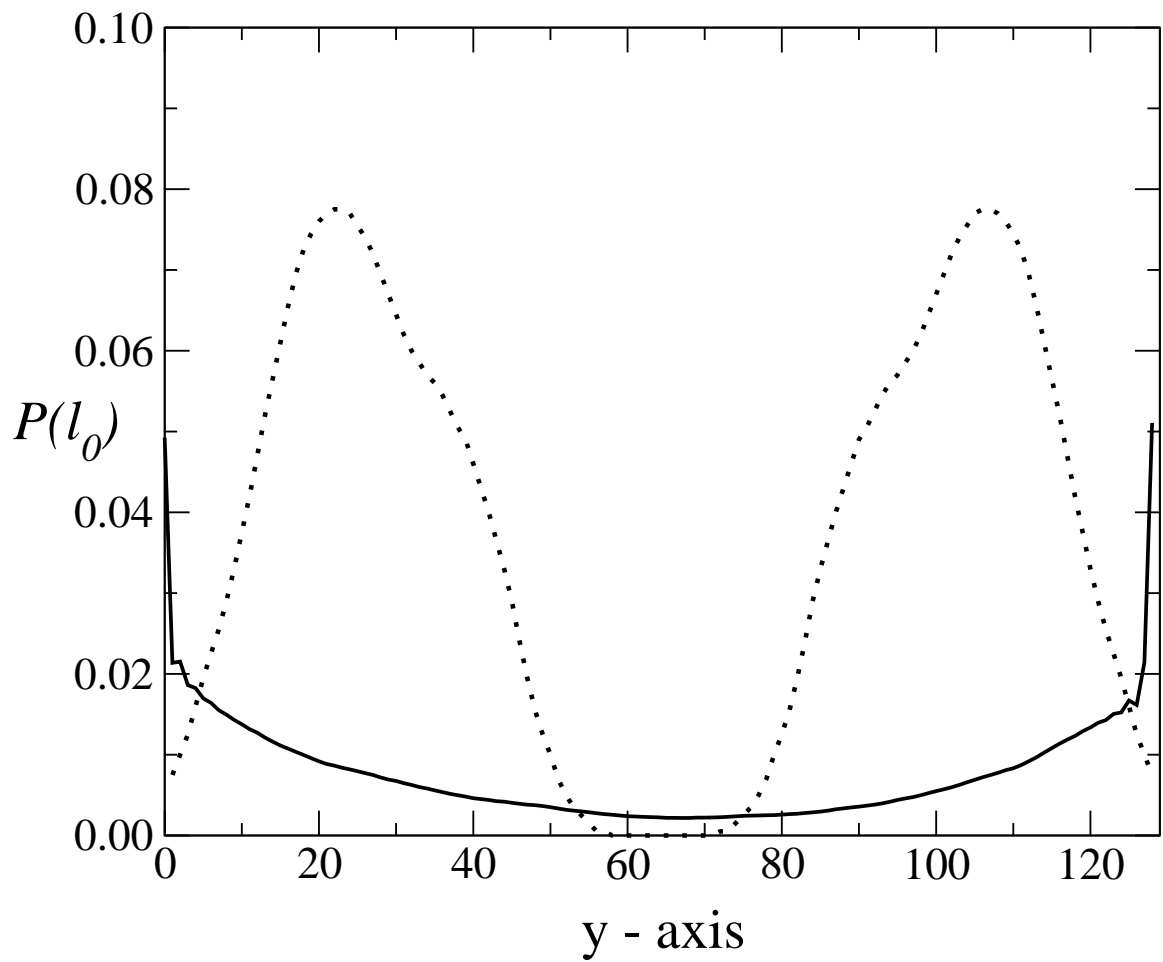}}}
\vskip 1.0 true cm
\caption{Probability distribution of the interface position $P(l_0)$ (full line) and damage $P(l_0^D)$ (dotted line) along the y-direction. Data corresponding to $L=128$, $T=0.90 T_C$, and $h=0.22 < h_f(\infty)$.}
\label{fig9}
\end{figure}

\begin{figure}
\centerline{{\epsfysize=5in \epsffile{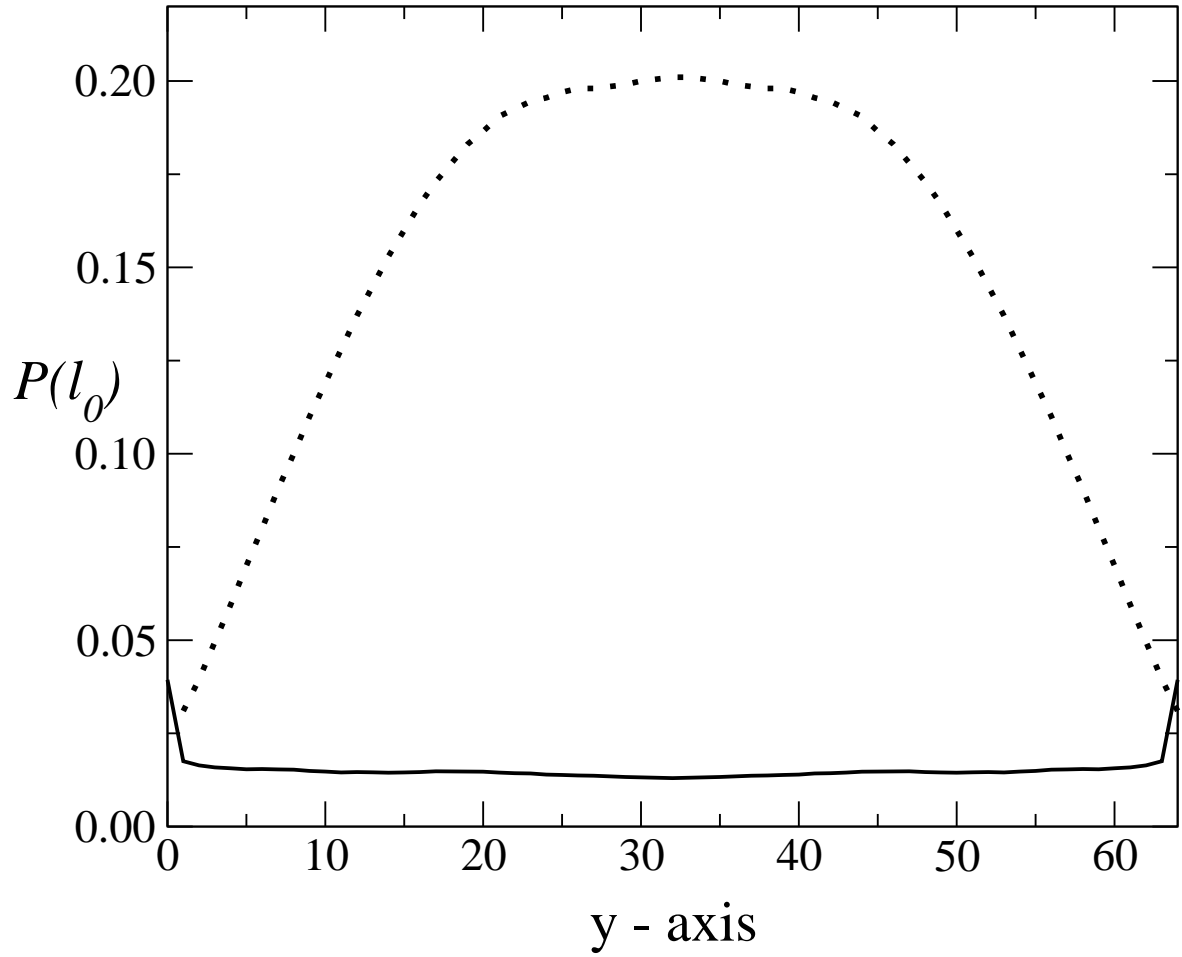}}}
\vskip 1.0 true cm
\caption{Probability distribution of the interface position $P(l_0)$ (full line) and damage $P(l_0^D)$ (dotted line) along the y-direction, for $L=64$, $T=0.90 T_C$, and $h=0.24 \approx h_f(\infty)$.}
\label{fig10}
\end{figure}

\begin{figure}
\centerline{{\epsfysize=5in \epsffile{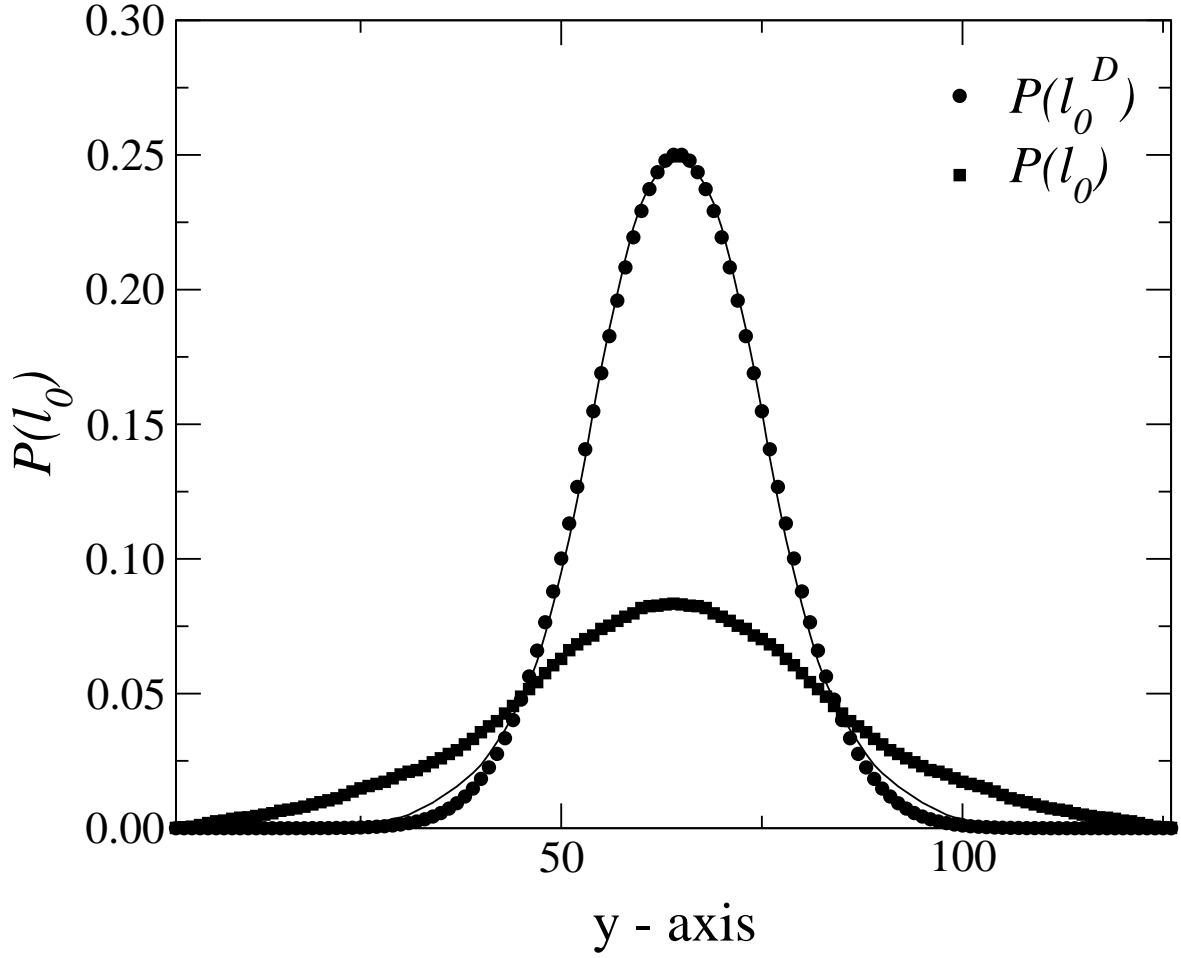}}}
\vskip 1.0 true cm
\caption{Probability distribution of the interface position $P(l_0)$ (squares) and damage $P(l_0^D)$ (circles) along the y-direction, for $L=128$, $T=0.90 T_C$, and $h=0.25 > h_f(\infty)$. The full line corresponds to the Gaussian fit of the data. The data corresponding to $P(l_0)$ were multiplied by a factor of 5 for the sake of clarity.}
\label{fig11}
\end{figure}

\begin{figure}
\centerline{{\epsfysize=5in \epsffile{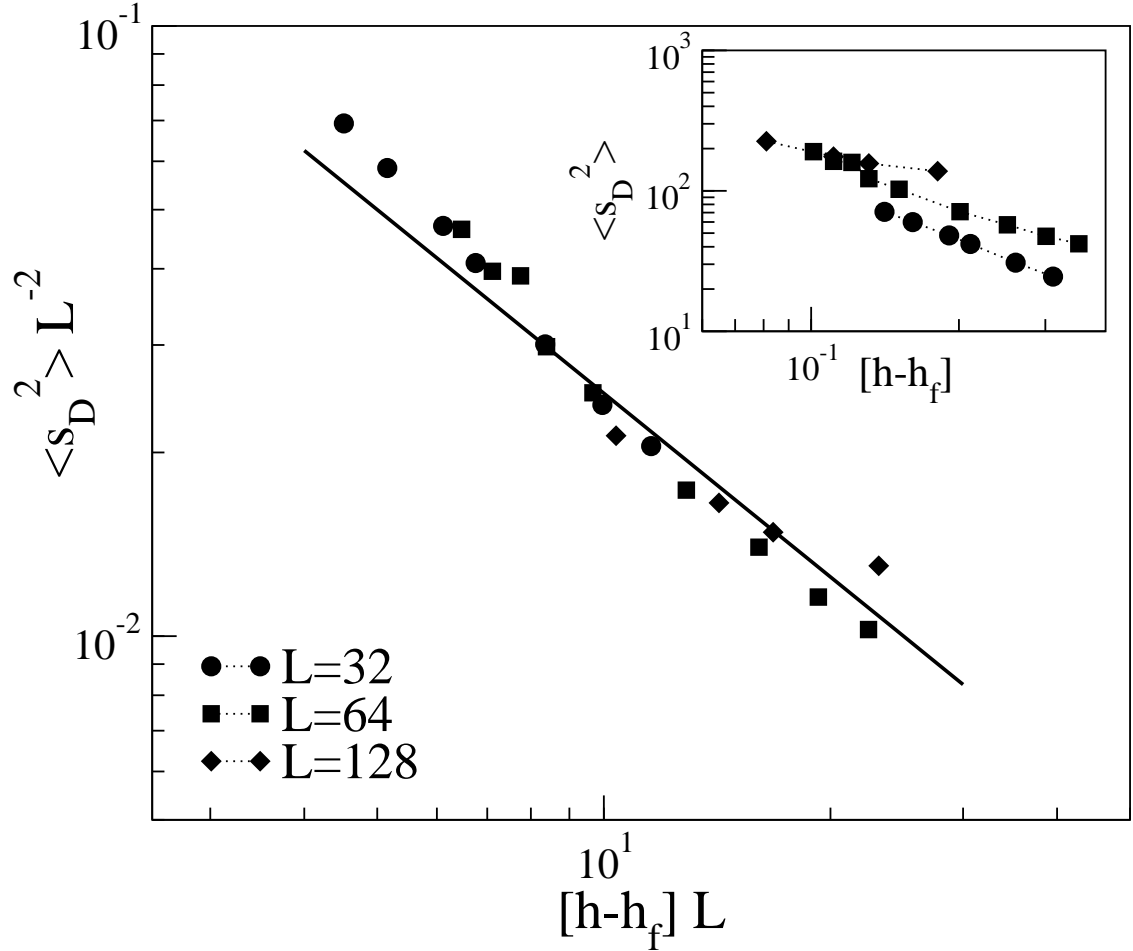}}}
\vskip 1.0 true cm
\caption{Width of the distribution of damage ($P(l_0^D)$) $\langle s^{2}_{D}\rangle$ versus $[h-h_{f}]$, obtained at $T=0.90 T_C$ and within the wet phase. Results corresponding to different values of the lattice size $L$, as indicated in the Figure. Inset: Collapse of the data according to equation (\ref{s2}) .}
\label{fig12}
\end{figure}

We studied the spreading of damage in the two-dimensional Ising Model confined in a corner geometry with a competing (short-range) magnetic field acting on the surfaces (see Figure \ref{fig2}).
By considering initial damage, created along one diagonal of the sample, in the direction perpendicular to the interface between domains of opposite magnetization originated by the applied fields, we conclude that the dynamics of damage spreading exhibits three characteristic regimes:
i) For short times, one observes the healing of the damage created in the bulk of the domains. 
However, at the critical damage spreading point, a small cluster of damaged sites always survives close to the interface.
ii) This small cluster propagates along the interface according to a power-law behavior $D(t) \propto t^{\eta^*}$, with $\eta^{*}=0.89(1)$.
By comparing the results of the present work and those already published \cite{alrp3} for the spreading of damage along the interface at the wetting transition, we conclude that the exponent $\eta^{I}=0.90(2)$ describes the critical propagation of the damage along magnetic interfaces.
iii) Finally, due to the constraint imposed by the corners where magnetic fields of opposite direction meet, the damage no longer propagates along the surface but starts to slowly spread into the bulk of the domains. 
Within this regime one has $D(t) \propto t^{\eta^**}$, where $\eta^{**}=0.40(2)$ is the exponent describing the spreading of the damage in the bulk.

We located the damage spreading transition by proper extrapolation to the thermodynamic limit. It is found that, within error bars, the phase diagram coincides with that of the corner filling transition. 
This result is in contrast to the previous study of the wetting transition where the damage spreading transition lies within the wet phase.
This difference can be understood on the basis of the constraint imposed by the field acting at the corners, causing a repulsive effect on the damage.
This repulsive effect is nicely observed in measurements of the probability distribution of the damaged area within the nonwet phase. 
Finally, in the wet phase the distribution of the damage is Gaussian and ist width scales as the fluctuations of the magnetization.

We hope that this study will contribute to the understanding of the propagation of magnetic perturbations along magnetic interfaces, as well as to the understanding of more complex damage spreading transitions that depart from the universality class of directed percolation.

\vskip 1.0 true cm
{\bf  ACKNOWLEDGMENTS}.
This work was supported financially by
CONICET, UNLP, and ANPCyT (Argentina). 
MLRP acknowledges CONICET for the grant of a fellowship.

\newpage

\newpage

\newpage

\newpage

\newpage

\newpage

\newpage

\newpage

\newpage

\newpage

\newpage

\newpage

\newpage


\begin{thebibliography}{}
\bibitem{dietrich} S. Dietrich, in {\it ``Phase Transitions and Critical Phenomena''}, Eds. C. Domb and J. L. Lebowitz, Vol. 12, Academic Press London (1988).

\bibitem{schick} M. Schick, in {\it``Liquids at Interfaces'' (Les Houches, Session XLV III)}, Eds. J. Charvolin, J. F. Joanny and J. Zinn-Justin, p 415, Amsterdam: Elsevier (1990).

\bibitem{abra1} D. B. Abraham, Phys. Rev. Lett. {\bf 44}, 1165 (1980).

\bibitem{abra2} D. B. Abraham, in {\it``Phase Transitions and Critical Phenomena''}, Eds. C. Domb and J. L. Lebowitz, Vol. 10 , Academic, New York (1987).

\bibitem{alba1} E. V. Albano, K. Binder, D. W. Heermann, W. Paul, Surf. Sci. {\bf 223}, 151 (1989); J. Chem. Phys {\bf 91}, 3700 (1989); Z. Physik B {\bf 77}, 445 (1989); J. Stat. Phys. {\bf 61}, 161 (1990).
 

\bibitem{macioleka} A. Maciolek, J. Stecki, Phys. Rev. B {\bf 54}, 1128 (1996). 

\bibitem{maciolekb} A. Maciolek, J. Phys. A {\bf 29}, 3837 (1996).


\bibitem{alba2} E. V. Albano, K. Binder, W. Paul, J. Phys. C {\bf 12} 2701 (2000).

\bibitem{binder2003} K. Binder, D. Landau, M. Miller, J. Stat. Phys. {\bf 110}, 1411 (2003).

\bibitem{duxb} P. M. Duxbury, A. C. Orrick, Phys. Rev. B {\bf 39}, 2944 (1989).

\bibitem{cheng} E. Cheng, M. W. Cole, Phys. Rev. B {\bf 41}, 9650 (1990).

\bibitem{napior} M. Napi\'{o}rkowsi, W. Koch, S. Dietrich, Phys. Rev. A {\bf 45}, 5760 (1992).

\bibitem{hauge} E. H. Hauge, Phys. Rev. A {\bf 46}, 4994 (1992).

\bibitem{lipow} A. Lipowski, Phys. Rev. E {\bf 58}, R1 (1998).

\bibitem{rejmer} K. Rejmer, S. Dietrich, M. Napi\'{o}rkowski, Phys. Rev. E {\bf 60}, 4027 (1999).

\bibitem{parry1} A. O. Parry, C. Rasc\'{o}n, A. J. Wood, Phys. Rev. Lett. {\bf 83}, 5535 (1999).

\bibitem{pwras} A. O. Parry, A. J. Wood, C. Rasc\'{o}n, J. Phys. C {\bf 12}, 7671 (2000).

\bibitem{glei} M. Gleiche, L. F. Chui, H. Fuchs, Nature {\bf 403}, 173 (2000).

\bibitem{rascon} C. Rasc\'{o}n, A. O. Parry, Nature {\bf 407}, 986 (2000).

\bibitem{parry2} A. O. Parry, C. Rasc\'{o}n, A. J. Wood, Phys. Rev. Lett. {\bf 85}, 345 (2000).

\bibitem{bed} A. Bednorz,  M. Napi\'{o}rkowsi, Phys. Rev. E {\bf 63}, 031602 (2001); J. Phys. A {\bf 33}, L353 (2001).

\bibitem{parry3} A. O. Parry, A. J. Wood, E. Carlon, A. Drzewi\'{n}ski, Phys. Rev. Lett. {\bf 87}, 196103 (2001).

\bibitem{parry4} A. O. Parry, A. J. Wood, C. Rasc\'{o}n, J. Phys. C {\bf 13}, 4591 (2001).

\bibitem{abra3} D. B. Abraham, A. O. Parry, A. J. Wood, Europhys. Lett. {\bf 60}, 106 (2002).

\bibitem{abraM}D. B. Abraham, A. Maciolek,  Phys. Rev. Lett. {\bf 89}, 286101 (2002).

\bibitem{parry5} A. O. Parry, M. J. Greenall, A. J. Wood, J. Phys. C {\bf 14}, 1169 (2002).

\bibitem{sart} A. Sartori, A. O. Parry, J. Phys. C {\bf 14}, L679 (2002).

\bibitem{abra4} D. B. Abraham, V. Mustonen, A. J. Wood, Europhys. Lett. {\bf 63}, 408 (2003).

\bibitem{albdevir} E. V. Albano, A. de Virgiliis, M. Müller, K. Binder, J. Phys. C  {\bf 15}, 333 (2003). 

\bibitem{milc} A. Milchev, M. Müller, K. Binder, D. P. Landau, Phys. Rev. Lett. {\bf 90}, 136101 (2003); Phys. Rev. E {\bf 68}, 031601 (2003).

\bibitem{parry6} A. O. Parry, J. M. Romero-Enrique, A. Lazarides, Phys. Rev. Lett.{\bf 93}, 086104 (2004). 

\bibitem{rom1} J. M. Romero-Enrique, A. O. Parry, M. J. Greenall, Phys. Rev. E {\bf 69}, 061604 (2004). 

\bibitem{rom2} J. M. Romero-Enrique, A. O. Parry, J. Phys. C {\bf 17}, S3487 (2005); Europhys. Lett. {\bf 72}, 1004 (2005).

\bibitem{abra5} D. B. Abraham, A. Maciolek, Phys. Rev. E {\bf 72}, 031601 (2005).

\bibitem{rejmer2} K. Rejmer, Phys. Rev. E {\bf 71}, 011605 (2005).

\bibitem{rascon2} C. Rasc\'{o}n, A. O. Parry, Phys. Rev. Lett. {\bf 94}, 096103 (2005).

\bibitem{giu} G. Giugliarelli, Phys. Rev. E {\bf 71}, 021603 (2005).

\bibitem{muller} M. M\"{u}ller, K. Binder, J. Phys. C {\bf 17}, S333 (2005).


\bibitem{manias} V. Manias, J. Candia and E. V. Albano, Eur. Phys. J. B. {\bf 47}, 563 (2005).


\bibitem{alrp1}M. L. Rubio Puzzo and E. V. Albano, Physica A {\bf 293}, 517 (2001).

\bibitem{alrp2}M. L. Rubio Puzzo and E. V. Albano, J. Magn. Magn. Mater. {\bf 241}, 110-117 (2002).

\bibitem{alrp3}M. L. Rubio Puzzo and E. V. Albano, Phys. Rev. B {\bf 66}, 104409 (2002).

\bibitem{herr90}H. J. Herrmann, Physica A {\bf 168}, 516 (1990).

\bibitem{herr92}H. J. Herrmann in {\it The Monte Carlo Method in Condensed Matter Physics}, K. Binder (Ed.), Springer, Berlin (1992).

\bibitem{note} Notice that for both the wetting and the corner filling transitions the magnetization is no longer the appropriate order parameter, as in the Ising Model, but the average location of the interface $\langle l_{0}\rangle$. So $\langle l_{0}\rangle \propto t^{-\beta_s}$ describes the divergence of the localized interface when the critical point is approached from below.

\bibitem{note2} Notice that $\langle l_{0}\rangle$ is measured in lattice units, so the total length of the lattice diagonal is $L$.

\bibitem{stan}H. E. Stanley, D. Stauffer, J. Kertesz, H. J. Herrmann, Phys. Rev. Lett. {\bf 59} 2326 (1987).


\end{thebibliography}
\end{document}